# Building AI and Human Capital for Road Safety

*Research-in-Progress Paper*

**Yug Dedhia**
International Institute of Information Technology, India
yug.dedhia@research.iiit.ac.in

**Anjali Singh**
International Institute of Information Technology, India
anjali.s@students.iiit.ac.in

**Vaibhav Singh Tomar**
International Institute of Information Technology, India
vaibhav.tomar@students.iiit.ac.in

**Nimmi Rangaswamy**
International Institute of Information Technology, India
nimmi.rangaswamy@iiit.ac.in

**Dev Singh Thakur**
i-Hub Data, International Institute of Information Technology, India
dev.singh@ihub-data.iiit.ac.in

## Abstract

AI is about learning algorithms and huge amounts of data and are drivers of economic growth- what does this mean for the field of development studies? Can we re-orient to twin AI studies and development theory and practice to generate how development challenges are identified and researched? To do this a good grasp is needed of AI's internal mechanisms and outcomes in addressing development issues – this argument will be developed through a case study of the ADAS [ Advanced Driver Assistance System] deployment in India. Over and above discussing the ADAS we bring an anthropological lens to understand the social context that surrounds the system. Focusing on bus drivers, we offer findings from a qualitative and ethnographic study of drivers in a collaborative effort to achieve road safety by deploying AI-driven technology and empowering stakeholders in the transport industry in India especially, bus drivers as critical actors in the city's transport network.

**Keywords:** Information Technology for Development (ICT4D), Human Capital, AI Capital, Anthropology, Road Safety, India

# Introduction

Driving is not merely a technical action. It is inherently social in which speed, distance, time is accounted for as part of the relations with drivers, vehicles, pedestrians, cyclists, among others, with whom we share a set of signals to guide traffic behaviors. Our primary research quest in this paper is to understand the advanced driver assistance system, henceforth ADAS, from the perspective of drivers and their interpretation of the outcomes of this system in benefiting everyday driving. India has one of the largest and the fastest growing road networks in the world making road safety a major developmental issue and public health concern. The Government of India is aiming for a 50% reduction in road fatalities on Indian roads by 2030 and boosting its road safety quotient with the predictive power of AI in enabling smarter and safer roads and drivers. iRASTE was officially launched in July 2022 by the Government of Telangana in South India, pushing AI driven advanced driver assistance solutions revolutionizing road safety (Chebbi, 2021). The AI-powered ADAS monitored the road ahead with a camera installed in each bus and via real time video analytics alerted the drivers of potential collisions with pedestrians, lane departures and forward collisions. The system prompted drivers to maintain safe distance from the vehicle ahead through real-time warnings and improve driver reaction time. Since periodic reinforcement of safe driving behaviors was essential, Project iRASTE will analyze driver performance by calculating the reduction in the number of alerts a driver received in a trip and a tangible reduction in accidents. Artificial Intelligence (AI) based solutions, in the current decade, have shown remarkable results worldwide in enabling safer Mobility and Transportation systems (Masello et al, 2022). Accordingly, the deployment of AI-based ADAS tools is viewed as a force multiplier in addressing problems of road safety on Indian roads. To accomplish the simultaneous building of both human and AI capital four major verticals were formulated around, vehicle safety, mobility safety, infrastructure safety and importantly, driver training and awareness to achieve the above three safety goals. Meeting all the above includes installation of ADAS devices on bus fleets for collection, storage, extraction, and collation of data on various alerts generated through the device. Based on the ADAS alert data AI capital is built via identifying grey spots on the road network followed by cost effective engineering improvements for possible implementation on these roads (Thakur et al, 2024). This is followed by the development of road quality index (RQI) through digital mapping of road infrastructure assets (Thakur et al, 2024). The bigger goal of the project is to generate rich and contextual information to build a technology ecosystem and nourish human capital that can be optimally operationalized by dynamically adapting to local road conditions.

The driver training and awareness critical in achieving enhanced road safety, is central to project iRASTE. Aligning with the holistic approach to road safety we study driver behaviors and their interactions with the technology system, particularly in the unstructured, non-delineated or well-defined Indian road infrastructures. An 'intelligent road' is associated with using, monitoring, and optimizing various digital technologies to maximize security, safety, and sustainability of road infrastructure. An 'Intelligent traffic monitoring system' will require enough information of how drivers of vehicles and road infrastructure could cooperate to improve overall traffic conditions- The question to ask her is - does research around intelligent road technology tend to focus in specialized areas without a holistic approach for deploying large-scale safety technologies? In this paper we discuss the move from the anxiety of building safe roads to focus on the problems related to the user/driver of the vehicle, the streams of information received and related dilemmas to support accurate, safe, and optimal decision making. For this task at hand, we need to know two very important things- first, is to establish the demands and requirements for the vehicle driver; Second, what should the vehicle know about the driver. The above offers unique opportunities for cross-disciplinary research between AI/ML system building and the social sciences towards achieving smarter effective road safety and to understand everyday driving practices to process, analyze and model data driven technology solutions.

We have two main research questions framing our study: One, to understand the ADAS from its beneficiaries, the bus drivers, and their assessment of the system; Two, to assess the integration of AI driven technology and human driven action on road safety in the specific context of iRASTE. Humans who interact with intelligent technologies develop an intimate understanding of the material world in which these technologies are deployed. AI technologies, especially those that are rendered 'autonomous', tend to subject perception of human action as passive recipients of technology prowess (Parvin & Pollock, 2020). A more broader research goal, we hope, is to mitigate this understanding through our research work in progress. We elaborate our methodology and discuss relevant prior research in the following sections.

## Methodology

We undertook several kinds of qualitative investigations, which is still in-progress: 1. A comprehensive tour of two bus depots in the city of Hyderabad and interviewing depot managers and engineers 2. Undertook five bus trips out of Hyderabad to study ADAS in real time, driver responses and its impact on driving culture. 3. We have attended several drivers training sessions since its inception from February 2023. 4. More importantly, we interviewed drivers one to one to develop social profiles and a holistic understanding of the driver- ADAS interaction and the driver response to technology driven road safety measures. Data from interviews and observational notes from bus trips were developed to focus on the actual use of, and interaction with, technologies in the wider context of people's lives and social structures. This approach (Horst et al , 2012) aids two processes: broad understanding of the wider work and social culture of research subjects, in this case the bus drivers, and more targeted research aimed at understanding a set of issues that arise from the driver-ADAS interaction. The focus is less on the technologies themselves than on their impact on users and outcomes thereof in specific locational or geographic contexts- in this instance the Indian roads. Our initial focus was to observe the processes by which drivers were responding to ADAS alerts and if they were changing behaviors in sync with the alerts. We used observational data during our bus trips to understand driver response and probed drivers to talk about the same while doing interviews. We broadly coded and organized data manually into themes and matrices shedding light on the specific aspects of driver ADAS interaction and response. More importantly, we explored how these unfolded in a context bearing technical and infrastructural constraints and in the broader social lives of drivers. Given our overall concern with how drivers understood and made sense of advanced assistance technologies, we focused on and addressed a) driver voices on ADAS b) the ways they respond to alerts, c) driver narratives on technology shaping their everyday work life d) broader social lives of drivers and their impact on work.

Our analysis of data was used to understand the following (we have not fully organized and developed insights as the research is still in progress) 1. Early driver response to the ADAS 2. Profiling drivers work culture and socio-economic background 3. Depot work practices in bus management and routines 4. Observing the ADAS system in real time from bus trips. Further, our second order analysis is currently in the process of exploring the following 1. To uncover anxieties, acceptance, and compliance in the driver-ADAS interaction mapping the relationship between driving experience and technology acceptance – for example does long distance driving bring more compliance than short distances? 3. Sketch and map infrastructural underpinnings for road safety and the corresponding outcomes for driver-ADAS interactions-example how does highways vs town/village road infrastructure impact driving? 4 A deeper understanding of driver social profiles and subjective voices on road safety, driving skills and technology.

## Literature Review

### *Building AI Capital*

There has been a spate of studies on road safety from urban India as well as other countries presenting innovative approaches to road safety management by integrating advanced technologies and methodologies. These studies also emphasize the potential for AI to enhance transportation infrastructure planning and align with the deployment of AI to develop smarter, safer, and more efficient transportation systems. Studies show the effective utilizations of artificial intelligence and data analysis to identify severely unsafe locations based on past fatalities, establishing a severity index, and enabling the prioritization of improvement strategies for these unsafe locations (Thakur et al, 2024) Advancement of the Internet of Things (IoT) and machine learning in the field of road safety and accident prevention builds technology capital via intelligent road safety systems. This study views the intersectionality of driver behaviors, the condition of vehicles, health condition of roads infrastructure among others to build an IoT enabled highly efficient road safety system updated on a real-time basis (Bhattacharya et al, 2022). A study of AIs potential as technology capital for road safety circumvents potential road danger through applications of convolution neural networks for image analysis (Ranganathan, 2022).

The safety potential of active and passive technologies in passenger cars in India uses data from the Road Accident Sampling System (RASSI) and develops rules to assess the effectiveness of technologies like Autonomous Emergency Braking (AEB) in preventing or mitigating accidents. The distribution of road traffic fatalities in India is analyzed, and an extrapolation method is used to estimate the benefits of technologies for the entire population (Puthan et al, 2018). Vehicles are increasingly equipped with sensors that capture the state of the driver, the vehicle, and the environment. State of a driver is often captured in data formats easily accessible as graphs of eye movements, headway, speed, or braking behavior. The study (Driessen, 2021) goes further to suggest that higher levels of decision making by the driver are context-dependent sources of information. Assessing a driver's abilities requires integrating multiple forms of 'knowing' about their driving and decision-making capacities in a state of stress and in the everyday routines of driving. This important study (Driessen, 2021) goes beyond simulator-based driver testing concluding that different forms of data are critical to evaluate drivers and driving abilities However, challenges exist in scaling accident databases and market penetration data for safety technology in India. The need for safety technology effectiveness in low and middle-income countries are always a collaboration between designing effective and applicable technologies that can complement and enhance human capacities to actively shape and contribute to public safety (Puthan et al, 2018)

### *Building Human Capital*

Building human capital through effective technology advancements are critical yet remain complex. We present a summary of research on human-technology vehicular eco systems and their outcomes- especially on drivers as primary stakeholders of public transport, and purported beneficiaries of advanced technology guiding systems. Addressing the pressing dangers of AI doesn't simply require attention to the data and models that inform computational action. It requires attention to the social structures that are built around AI technologies (Fox, 2023). Researchers go on to point to the criticality of anticipating all aspects of man-machine integration before a technology is implemented. Yet, many of the consequences of technology that are deemed "unanticipated" are oversights – resulting from design processes that fail to include a diversity of stakeholders (Parvin & Pollock, 2020). Research on the behavioral characteristics of drivers through qualitative studies like interviews, questionnaires, laboratory tests, simulation studies and field studies which help psychological behaviors while driving like rage enactment, sensation or thrill-seeking or attention depleting behaviors. Human behavioral studies simulating real world scenarios have pointed to over-reliance on assistive driving can lead to weakened driver input and potential accidents (Thakur et al, 2024). A systematic review was conducted to investigate the relationship between Advanced Driver Assistance Systems (ADAS) usage and driver distraction Drivers had longer reaction times to critical events despite reporting higher trust and perceived safety and decreased workload (Hungund et al, 2021). A study highlighting driving contexts and cultural factors found Swedish drivers were more likely to use ADAS than Chinese and American drivers due to several factors- differences in driving culture, traffic regulations, and vehicle technology (Orlovskaa et al.,2020). Another study finds 'experience' an important factor in adapting to and using driver-assistance systems safely and effectively. Experienced drivers adapted to the ADAS more quickly and felt more comfortable using them than inexperienced drivers. Experience likely aided drivers to constantly monitor their surroundings effectively (Ucińska, 2021). Advanced features can help to reduce driver errors and improve safety on the road (Pulugurtha et al, 2022) - for example, adaptive cruise control (ACC), lane departure warning (LDW), automated emergency braking (AEB), and impacted driving behaviors towards better safety. However, drivers needed heightened awareness of technology limitations and using them responsibly. Also, unfamiliarity with advanced features likely abetted misuse or rendering the driver more complacent. The importance of effective technology-driver interactions to understand advanced features needed to be combined with safe driving practices and driver's control of the vehicle may not be substituted for technology aides (Pulugurtha et al, 2022)

## Findings and Discussion

AI is about learning algorithms and huge amounts of data as drivers of technology innovation and growth- what does this mean for the field of development studies. Can we re-orient and twin AI studies and development theory and practice to generate how development challenges are identified and researched? To do this we need a good grasp of AI's internal mechanisms and outcomes in addressing development issues and we do this through a case study of the ADAS deployment in India in project iRASTE. Over and above discussing the ADAS we bring a social lens to understand contexts surrounding the technology system. Focusing on bus drivers, we offer findings from a qualitative and ethnographic study of drivers in a collaborative effort to achieve road safety deploying AI driven technology and empowering stakeholders in the transport industry in India especially, bus drivers as critical actors in the city's transport network. Our study is an engagement to understand the social contexts of and their consequences for bus drivers to adopt, reject or remain indifferent to technology support. We developed insights from our sociological data analysis, mostly the drivers view of the ADAS and consequent appraisals of driving behaviors towards better road safety 1. The ADAS system triggers mixed early reactions among drivers 2. Drivers are affected by their broader socio-economic lives that impact technology adoption 3. The ADAS alerts need not necessarily reflect driver performance 4. It is critical to understand and /or evaluate driver performance (without ADAS) before we rank them through ADAS.

### *Driver Background*

Several drivers, more than half of our interview pool (while we write this paper, we had interviewed around 32 drivers) had 20 years of driving experience- one of them has been driving for 40 years. Most had 10 years of driving experience- a few of them had past jobs driving trucks, some were bus conductors. One of the drivers had been driving for a national airline inside of the premises of a mid-sized airport. Many had learnt driving hands-on in their hometowns and procured formal driving licenses (we have not yet discussed their early motivations to become professional bus drivers). Our data analysis focused on first person narrations of the driver's social and work lives. We seek to develop, from these narrations, a driver perspective about the nature, challenges, and responses to technology support at work. It is not surprising to see participants associating the personal and the social with work experiences. As one of the authors of this paper is trained in social anthropology, the study integrates the driver point of view about the ADAS as part of the everyday professional and personal work experience.

### *The Work Context*

There were two sets of drivers- one set were working for the government and are formal employees of the government of Telangana with mandated employee benefits (health insurance, saving schemes among others); the second set were under a private contract to provide transport services to the government. employees of the state government entitled to all. The second set of drivers are conscious of their makeshift work status (despite serving a government body) and the resulting lack of agency in negotiating pay structures, work conditions, and more importantly, the burden of responsibility in the rendering of their public-facing services. In our discussions, this lack of agency constantly permeated any talk about ADAS or road safety measures – road safety for the bus driver is not only about the act of driving- it is about the role they play in the transport system as a key stakeholder and service provider. A driver (with a 10-year experience) told us, "… we are major arbitrators of road safety but are not treated as one- we take pay cuts when vehicles are damaged, sometimes unfairly… I had to pay up for bumping and damaging a vehicle (in front) when the brake in my bus failed! Is it reasonable to expect us to upkeep vehicles- these new technologies will also need upkeeps/upgrades- are we expected to do this too…"

### *Driving as a Social State*

"Do we as drivers, leave our personal problems at home and focus on the road? Can we come with a clean slate to drive each day?" asked a senior driver of more than 30 years of driving. He explained that driving, to him, is first and foremost a state of mind that demands attention, focus and commitment, which must be achieved only through the perseverance of the driver. Most drivers are proficient and constantly adjusting to the dynamics of urban road conditions. They face poor roads, pedestrian unruliness, traffic indiscipline and constant road re-constructions – 'apane hisaab se chalate hain log' voiced by many drivers who navigate the uncertainties of urban road fabric and location configurations. Many spoke about endeavors to save life and property through conscientious driving. It becomes interesting to view technology as an adjudicator of driver performance- the ADAS system ranking drivers based on the volume/instances of CAS alerts.

### *ADAS, a Socio-Technical System*

Drivers are enrolled in a ranking system drawn from the number of alerts (four types of alerts) eventuating from the Collision Avoidance System installed in buses (at the time of this study there were a total of 100 buses with ADAS installation and around 250 drivers driving them in shifts). A drop in the volume of CAS alerts advance the rank of a driver over a period and is incentivized via small gifts to the driver maintaining or advancing towards a higher rank.

### *The Driver Point of view*

Most drivers experience a lack of control over alerts that do not concern (bad)driving behaviors 1. stopping for red light the bus moves very close to the vehicle ahead and triggers an alert before the bus comes to a halt- Mohammed, a driver with 5 years of experience said, " we cannot leave even a one-feet gap from the vehicle in front- someone is going to but in-I need to halt close to the vehicle in front and the buzzer( CAS alert) is triggered…" 2. Vehicles that overtake from the side and drive close to the bus to get ahead/out front thus triggering an alert 3. bumper to bumper traffic always produces alerts until the bus comes to halt (despite the halting of CAS when the speed of the bus touches zero) 4. adjusting for rogue road discipline, heavy traffic caused by construction work, the problem of two wheelers and the new electric motor vehicles (that stop anywhere to accept commuters), roadblocks, crowded spaces 5. cautious driving behaviors resulting in overshooting bus time schedules, a driver with 4-year experience told us, "… there is a new rule that got implemented whereby we need to stop at every bus stand for 2 seconds even if there are no commuters waiting; this messes up with our time/schedules … we sometimes ignore alerts buzzing from the system as we are unable to slow down or exercise caution to maintain bus schedules 5. Work shifts leading to lack of control over vehicle upkeep/upgrades, as Mohammed told us "Bus/vehicles are rotated, and I am never driving the same vehicle for more than 2 months at a stretch- how will I have any control over its condition? How can I be responsible for the vehicle's quality?"

### *Building Driver Performance and Work Capital*

In the time spent with drivers, we sensed an overall lack of engagement with the ranking and reward system. Most drivers were tending towards an acceptance of ADAS as a 1. Record keeper of driving behaviors 2. An alert system to avoid accidents 3. A provider of data on road conditions, accidents, or events. There was an overwhelming acknowledgement of positive impacts via CAS averting accidents- one of them talked about saving a cow and another saving a pedestrian from an accident, though minor, only due to the CAS alert. We attempted to probe if CAS alerts generated more focus and concentration in the long term. There was no discussion on the long-term transformations in driving behaviors the ADAS can potentially bring about. Drivers offered a few suggestions to advance the ADAS 1. better camera coverage for alerts especially the breath of vision and side angle vision 2. a longer alert time, more than 2 seconds though we did discuss collision avoidance as a time constrained alert, a function of distance and cannot be predicated ahead of time. Drivers spoke about these suggestions as impacting driver performance as evaluated by ADAS.

### *Evaluating Driver Performance*

This section offers reflections of driver performance as an evaluation of the ADAS. Drivers spoke with pride about skills they develop on the job. Many recounted never having been in accidents, to life or property, in their driving careers. They constantly drew our attention to the driving skills that are built from both experience and instinct and together produce a wisdom to navigate/mediate traffic conditions. How is the ranking system taking account of the proficiency of drivers that evolves over time and experience? Machine learning, recognition, and intelligence, through CAS, constitute ranking of driving behaviors based on apriori definition of cautious and safe driving practice. How do we account for the experience of drivers that has already rendered them as skilled workers? Is it fair to judge drivers through a narrow definition of safe driving mediated through a technology driven support system? Another question is the understanding of the ADAS/CAS as influencing drivers to achieve better driving performance or simply recording good drivers who maintain or advance in rank. Who is a good driver and how are we making this comparison? What are the metrics, reference points, that are not defined by ADAS? Do we have data that can compare drivers under similar conditions!  Does the CAS alert data tell us about the condition of the bus or the road or the nature of pedestrian or traffic conditions when a collision is averted or forewarned due to driver behaviors? We are reminded of the following quote from a driver we quoted in the beginning of this report "…I had to pay up for damaging a vehicle in front of me when the brake in my bus failed! Can the CAS alert offer data that tells us it's not the driver but the bus condition that caused the accident…". Several drivers viewed alerts ensuing from CAS as a 'mistake' or breaking a rule of safe driving.  A better understanding of ADAS is needed for drivers to respond favorably and make a concerted effort to interact with the technology system. In our understanding, drivers who consistently receive low ranks are not followed up with incentives or even serious appraisal. If we might add, it's tradition in anthropological research to offer regular feedback to participants/subjects of research, especially when they are part of an experiment - in this case the ADAS performance and ranking scheme needs informed participation from drivers, more importantly their buy-in to enroll and engage with the system.

*Final Thoughts*

The focus on drivers as recipients of the ADAS and undertaking a qualitative field study to develop and understand our research goals have yielded a specific perspective on the assessment of AI capital to aid and assist in building human capital. The translation of AI capital to human capital is not a straightforward path – there are multiple conditions, both human and infrastructural, that impinge on the collective potential of human and AI to aid road safety. We argue the social science perspective and methodologies offer a fruitful way to unpack the integration of AI-Human capital towards a techno-social outcome. In this paper we offer ways to study the coming together of technology and people going beyond machine data by including social data. We hope to have opened a research channel to understand real world human interactions with technology, particularly in impacting social benefits for society at large.

To conclude we would like to raise a few thoughts on building human capital through AI technologies and current capabilities of the ADAS as a driver incentivizing mechanism for road safety behaviors. Data from ADAS aiding the assessment of driver performance for road safety also collects and detects gray spots, road conditions, location coordinates and road architectures/configurations to get a deeper understanding of road risk and safety. Can this data offer intelligent interplays of driver performance and matching data about road conditions and their location? The latter can address driver concerns about the lack of control in reducing ADAS alerts and bearing singular responsibility for unsafe driving practices. Integrating data about driver history and driving practices while ranking them through ADAS can offer a holistic and fair understanding of driving skills and drawbacks. Driver appraisal and feedback need serious engagement to build human capital and render drivers as critical stakeholders of road safety practices. Our paper is an attempt to bridge cross-disciplinary research between AI/ML system building and the social sciences to analyze and model data driven technology solutions. It is also an effort to bring a holistic approach for deploying large-scale intelligent road technology by adding a human focus, moving away from a singular focus on technology and bringing problems related to the users of technology – in this paper we believe to have added valuable data on human stakeholders, their social contexts and ecosystems surrounding AI driven technologies. Future work aims to integrate theories from development studies, such as the capabilities framework, integrating broader human stakeholders in the research of socio-technical systems and offering stronger research embeddings in the domain of ICTD.

## Acknowledgements

We are grateful to iHub- Data, IIIT Hyderabad for the research grant to explore project iRaste. This paper is an outcome of this research. We thank Sridhar M who was integral to our research process.